\documentclass[
    ,final            
  ]
  {aipproc}

\layoutstyle{8x11double}

\renewcommand\XFMtitleblock{%
  \XFMtitle
  \let\XFMoldpar\par
  \def\par{\XFMoldpar\def\par{\space
             (on behalf of the CANGAROO-III Collaboration)\XFMoldpar}}%
   \XFMauthors
   \let\par\XFMoldpar
   \XFMaddresses
   \XFMabstract
   \vspace{5pt}%
   \XFMkeywords
   \XFMclassification
 }

\begin{document}

\title{The implications from CANGAROO-III observations of TeV blazar 
PKS\,2155-304}


\author{K. Nishijima}{
  address={Department of Physics, Tokai University, Hiratsuka, Kanagawa
  259-1292, Japan}
}

\author{Y. Sakamoto}{
  address={Department of Physics, Tokai University, Hiratsuka, Kanagawa
  259-1292, Japan}
}

\author{J. Kushida}{
  address={Department of Physics, Tokai University, Hiratsuka, Kanagawa
  259-1292, Japan}
}

\author{K. Saito}{
  address={Department of Physics, Tokai University, Hiratsuka, Kanagawa
  259-1292, Japan}
}

\begin{abstract}
 We have observed the high-frequency--peaked BL Lacertae 
object PKS\,2155$-$304 in 2004, 2005 and 2006 with the CANGAROO-III 
imaging atmospheric Cherenkov telescope, and have detected 
a signal above 660~GeV at the 4.8~$\sigma$ level during 
the 2006 outburst period. 
Intranight flux variability on time scale of half an hour
is observed. From this variability time scale, 
the size of the TeV gamma-ray emission region
is restricted to $5\times10^{13}\delta$~cm, and
the super massive black hole mass is estimated to
be less than $1.9\times10^8{\delta}\rm M_{Solar}$,
where $\delta$ is the beaming factor.
The differential energy spectrum is obtained,
and an upper limit of the extragalactic infrared background 
light (EBL) flux is derived under some assumption. 
We also fit a synchrotron self Compton (SSC) model 
to the spectral energy distribution (SED) and derive
the beaming factor and magnetic field strength. 

\end{abstract}

\maketitle


\section{Introduction}

The first detection of TeV gamma rays from PKS\,2155$-$304 
was reported by the Durham group during the active phase 
in 1997\,\citep{Cha99}.
Although the CANGAROO-I and the CANGAROO-II observations 
in 1997, 1999, 2000, and 2001 could not find an excess signal
from PKS\,2155$-$304\,\citep{Rob99,Nis01,Nis02,Nak03}, 
it was confirmed by the H.E.S.S.\ group as
a TeV gamma-ray source in 2004\,\citep{Aha05}. 
In July 2006, the extreme TeV flare was detected by 
the H.E.S.S. group, and they reported very beautiful 
light curve with 1 minute timescale resolution\,\citep{Aha07}.
Some modeling for the rapid TeV flux variability and 
the spectral energy distribution of PKS\,2155$-$304 
are reported\,\citep{Fos07,Ghi08,Kus08,Kat08}, 
and intensive interpretations of the data have been 
attempted.

CANGAROO-III observations of PKS\,2155$-$304 have been performed
from 2004 to 2006. Particularly in 2006, target-of-opportunity
observations were made triggered by the H.E.S.S. report.
In this paper, we summarize our observations and 
the results particularly focused on the data of the 2006 
outburst period\,\citep{Sak08}, and give some implications from them.

\section{Observations and Analysis}

The CANGAROO-III imaging atmospheric Cherenkov telescope system
is operated in Woomera, South Australia (longitude 
$136^{\circ}47^{\prime}$E,
latitude $31^{\circ}06^{\prime}$S, 160~m a.s.l.).
Three of the four telescopes were used in these observations.
The details of the CANGAROO-III telescope system are
described in \citep{Kaw01,Kab03,Eno06a}.

Observations of PKS\,2155$-$304 in 2004 were made 
by local trigger using three telescopes. 
From 2005, an on-line stereo trigger system
became available.
The details of the CANGAROO-III trigger system are
given in \citep{Kub01,Nis05}.
In 2005, observations were made with only two telescopes due to 
electronics problems.
Observations in 2006 consist mostly a three-telescope configuration,
but the exception to this is in a part of July due to a mechanical
tracking problem with the third telescope.
These observations were made using wobble mode,
in which the pointing position of each telescope was shifted 
in declination by $\pm0.5^{\circ}$ from the center of PKS\,2155$-$304
alternatively every 20 minutes.
Observation periods, observation nights, the amount of observation time,
the number of telescopes available, and the trigger mode 
for each year are summarized in Table\,\ref{tab:summary_obs}.
Typical trigger rate of the two-fold and the three-fold coincidences
on July 2006 observations were $\sim20$~Hz and $\sim12$~Hz,
respectively.

After the image cleaning and the reconstruction of 
the arrival direction,
we applied the Fisher discriminant~(FD) method to the data 
in order to reject numerous cosmic-ray background events\citep{Eno06b}.
Following a Monte Carlo simulation study, the optimum FD
cut values were determined, and the $\theta^2$ cuts were
applied, for example at $\theta^2 < 0.06~\mbox{deg}^2$
for the three fold data. Details of the observations 
in 2006 and analysis method are described 
in \citep{Sak08}, and references therein.  
After the data reduction described above, 
the effective live times are
17.0~hrs, 38.6~hrs, 25.1~hrs, and 17.1~hrs in 2004,
2005, July 2006, and August 2006, respectively.

\begin{table}
\begin{tabular}{lcrccc}
\hline
  \tablehead{1}{l}{b}{Year}
  & \tablehead{1}{c}{b}{Period}
  & \tablehead{1}{r}{b}{Nights}
  & \tablehead{1}{c}{b}{Obs. time~[hrs]}
  & \tablehead{1}{c}{b}{$\sharp$ of tel.}  
  & \tablehead{1}{c}{b}{Stereo mode}   \\
\hline
2004 & Aug.8-23 & 11 & 20.5  & 3 & off-line\\
2005 & June 6-15 & 6 & 46.8  & 2 & on-line\\
     & July 1-5 & 5 & $\uparrow$  & 2 & on-line\\
     & July 29-Aug.5 & 12 & $\uparrow$  & 2 & on-line\\
2006 & July 28-Aug.2 & 5 & 29.0  & 3\&2 & on-line\\
     & Aug.17-25 & 6 & 19.1  & 3 & on-line\\
\hline
\end{tabular}
\caption{Summary of CANGAROO-III observations
of PKS\,2155$-$304}
\label{tab:summary_obs}
\end{table}

\section{Results}

\begin{figure}
  \includegraphics[height=9pc]{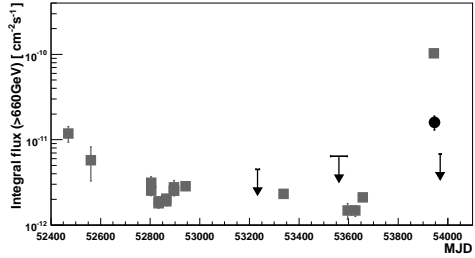}
  \caption{Long-term variations of the flux of
PKS\,2155-304~(black closed circle and bars with an arrow) 
compared to H.E.S.S. results~(gray squares)\citep{Aha07,Pun07}.
The flux of the H.E.S.S. are converted to the values
above 660~GeV assuming a photon index of 3.3. }
\label{fig:allyear}
\end{figure}

In 2004 and 2005, no significant excess events were found and
$2~\sigma$ upper limits on the integral flux were obtained:
$4.5\times10^{-12}\mbox{cm}^{-2}\mbox{s}^{-1}$
above 580~GeV for 2004 and 
$6.4\times10^{-12}\mbox{cm}^{-2}\mbox{s}^{-1}$
above 660~GeV for 2005.
In July 2006, just after the largest outburst reported by 
the H.E.S.S.\ group\citep{Ben06},
we detected $322\pm67$ excess events corresponding to 
$4.8 \sigma$ level from the direction of PKS\,2155$-$304\citep{Sak08}.
The time-averaged integral flux above 660~GeV is
calculated to be
$F(>660~\mbox{GeV})=(1.6\pm0.3_{stat}\pm0.5_{syst})\times10^{-11}\mbox{cm}^{-2}\mbox{s}^{-1}$,
which corresponds to $\sim45\%$ of the flux observed
from the Crab Nebula\,\citep{Aha04}.
In the follow-up observations in August 2006, 
only the flux upper limit ($2~\sigma$ level) was
obtained:
$F(>660~\mbox{GeV})<6.4\times10^{-12}\mbox{cm}^{-2}\mbox{s}^{-1}$
(20\%~Crab flux), which means the
TeV gamma-ray activity subsided in August.
These results are summarized in Table\,\ref{tab:summary_flux}.
Our results obtained with the CANGAROO-III telescope for each year 
are compared to the H.E.S.S.\ results
in Fig.\,\ref{fig:allyear},
where the fluxes by the H.E.S.S. are converted to the values
above 660~GeV assuming a photon index of 3.3 and plotted.

\begin{table}
\begin{tabular}{lrccccc}
\hline
  \tablehead{1}{l}{b}{Year}
  &
  & \tablehead{1}{c}{b}{$E_{th}$ \\\ [GeV]}
  & \tablehead{1}{r}{b}{Excess \\events}
  & \tablehead{1}{c}{b}{Significance \\\ [$\sigma$]}
  & \tablehead{1}{c}{b}{Flux \\\ [$10^{-12}\mbox{cm}^{-2}\mbox{s}^{-1}$]}  
  & \tablehead{1}{c}{b}{Crab flux\\\ [$\%$]}   \\
\hline
2004 & & 580 & -14$\pm$28 & -0.5  & <4.5 & <10\\
2005 & & 660 & 64$\pm$53 & 1.2  & <6.4 & <20\\
2006 & July & 660 & 322$\pm$67 & 4.8  & 16$\pm3_{stat}\pm5_{syst}$ & 45\\
&     Aug. & 660 & -1$\pm$30 & -0.0  & <6.5 & <20\\
\hline
\end{tabular}
\caption{Summary of results for PKS\.2155-304, where
2006 July corresponds to period from July 28 to August 2 2006}
\label{tab:summary_flux}
\end{table}

\begin{figure}
  \includegraphics[height=10pc]{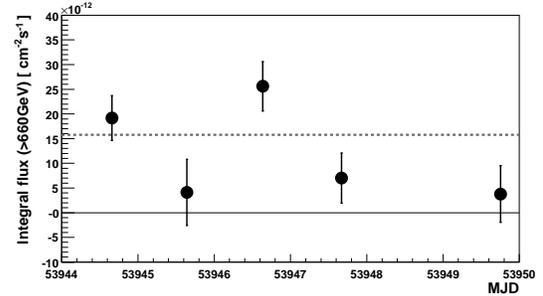}
  \caption{Daily light curve of PKS\,2155-304 from July 28 to
August 2 in 2006 expressed by the integral flux above 660~GeV.
The dashed line indicates an average integral flux during this 
observation period.}
\label{fig:light_curve_July}
\end{figure}

\begin{figure}
  \includegraphics[height=10pc]{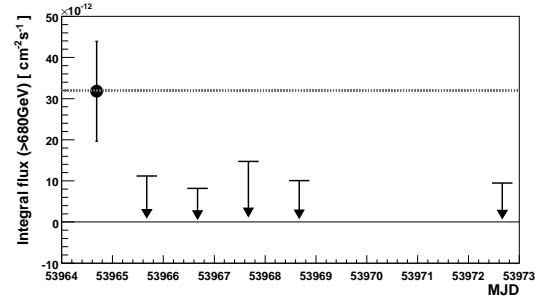}
  \caption{Daily light curve of PKS\,2155-304 from August 17 to
25 in 2006 expressed by the integral flux above 660~GeV.
The dotted line indicates one Crab flux level above 660~GeV.}
\label{fig:light_curve_Aug}
\end{figure}

Fig.\ref{fig:light_curve_July} and \ref{fig:light_curve_Aug} 
show the nightly average integral flux above 660\,GeV
in 2006. From the light curve, it is seen that 
the averaged flux reached 
$\sim70\%$ of Crab flux level on the night of July 30.
Assuming a constant average flux between July 28 and August 2,
a $\chi^2$ fit yields a value of 13.9 for 4 degree of freedom,
which corresponds to a $\chi^2$ probability of $\sim8\times10^{-3}$.

\begin{figure}
  \includegraphics[height=9pc]{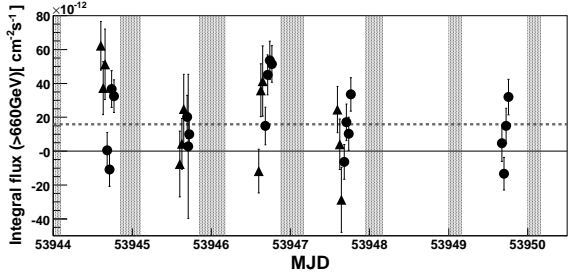}
  \caption{Light curve of PKS\,2155-304 expressed by the
integral flux above 660~GeV between July 28 and August 2
based on a 40 minute bin width. Triangles and closed circles
indicate the flux obtained from the two-fold and three-fold
data set, respectively.The dashed line shows an average integral
flux during this observation period. The shaded areas indicate
the H.E.S.S. observation periods. The time difference between
the H.E.S.S. and CANGAROO-III sites is $\sim8$ hours.}
\label{fig:light_curve_40min}
\end{figure}

We further investigate intranight variation by dividing
the data into $\sim40$ minute bins for each night in July 2006. 
The light curve is shown in Fig.\,\ref{fig:light_curve_40min}. 
Assuming a constant average flux for each night, 
$\chi^2$ test for the data of July 28 and 30 gives
$\chi^2/dof=29.2/6$ and $22.1/6$, respectively, 
which correspond to $\chi^2$ probability of $6\times10^{-5}$
and $1\times10^{-3}$, respectively. The intranight variations
are apparent.
The fractional root mean square variability amplitudes
$F_{var}$ are calculated to be $0.75\pm0.07$ on July 28 and 
$0.58\pm0.08$ on July 30.
These values are comparable to the variability reported 
by the H.E.S.S.\,\citep{Aha07}.
Doubling times are also calculated from Fig.\,\ref{fig:light_curve_40min}. 
We obtained the shortest doubling time of 34 minutes
as an $1\sigma$ upper limit.

A photon index of the time-averaged
differential energy spectrum of PKS\,2155$-$304, 
which was obtained between July 28 and August 2 in 2006,
is $\Gamma=-2.5\pm0.5_{stat}\pm0.7_{syst}$,
and a flux normalization at 1 TeV is
$N(1~\mbox{TeV})=(1.0\pm0.2\pm_{stat}\pm_{syst})\times10^{-11}\mbox{cm}^{-2}\mbox{s}^{-1}\mbox{TeV}^{-1}$\citep{Sak08}.

\section{Discussion}

Assuming that the gamma-rays from PKS\,2155$-$304 are
generated in the jet which is directed toward us with 
a beaming factor $\delta$, the
observed variability time scale $t_{var}$ is related to 
the size $R$ of emission region 
by the equation $R<{\delta}ct_{var}/(1+z)$ from causality, 
where $z$ is the redshift.
Using an upper limit of the shortest doubling time 
as an variability time scale, $i.e.$ $t_{var}=34$~minutes,
the size of TeV gamma ray emission 
region is limited to $R<5\times10^{13}\delta$~cm.
The central engine of a blazar is believed to contain 
a super massive black hole.
If we assume that the size of emission region
is larger than the Schwarzschild radius $R_{Sch}$, $i.e.$ $R>R_{Sch}$, 
a black hole mass $M_B$ is less than $(c^2/2G)R$.
Then, substituting numerical values of constant, 
an upper limit of a  super massive black hole mass 
is estimated to be
$M_B<1.9\times10^8{\delta}\rm M_{Solar}$, 
where $\rm M_{Solar}$ is one Solar mass.
On the other hand, assuming an internal shock model in the jet, 
where the second blob catching up with the first blob and 
create a shock wave,
we estimate the distance $D$ from the center of black hole to 
the shock region as
$D=10^3(k/10)(\gamma/10)^2[2\alpha^2/(\alpha^2-1)]R_{Sch}$.
Here $\gamma$ and $\alpha\gamma$ are Lorentz factors of the first 
and the second blobs($\alpha>1$), respectively, 
and blobs are assumed to be emitted at a time interval
of ${\sim}kR_{Sch}/c~(k>3)$.
The size of gamma-ray emission region is $D/\gamma$
and observed variability time scale is $t_{var}{\sim}D/c\gamma^2$.
So the black hole mass is expressed as
\begin{displaymath}
M=1.0\times10^6 \left(\frac{t_{var}}{sec}\right)
\left(\frac{10}{k}\right)
\left(\frac{\alpha^2-1}{2\alpha}\right)\rm M_{Solar}.
\end{displaymath}
Substituting $t_{var}=34$ minutes, and assuming $\alpha\gg1$ and $k>3$,
the black hole mass is estimated to be less than $3.4\times10^9\rm M_{Solar}$.

As is well known, TeV photons are absorbed by an interaction
with extragalactic infrared background light~(EBL) through
the pair production process 
$\gamma_{TeV}+\gamma_{EBL}{\rightarrow}e^++e^-$.
Measured flux $F_m$ is related to the source flux by
$F_m=e^{-\tau}F_s$, where $\tau$ is the optical depth
of the above process. 
Assuming an intrinsic photon index is not harder than $-1.5$\citep{Mal01}
and no absorption below 200\,GeV, the upper limit of 
the optical depth $\tau$ is calculated from 
the observed spectrum.
Adopting the $1\sigma$ upper limit of the observed spectrum 
in July 2006, the optical depth, for example at $1.1{\mu}m$, 
is calculated to $\tau=2.3$, and  
an upper limit on the EBL flux
can be estimated to be $45.5 \mbox{nWm}^{-2}\mbox{sr}^{-1}$
for $1.1{\mu}m$.

On the contrary, assuming the EBL density model, an intrinsic
spectrum can be estimated from the measured spectrum.
Choosing the EBL density model of Primack et al.\citep{Pri05},
an intrinsic spectrum of 
$dN/dE=(2.7\pm0.6)\times10^{-11}(E/1\,TeV)^{-(1.8\pm0.6)}cm^{-2}s^{-1}TeV^{-1}$
is derived from the observed spectrum. 
When applying the model by Stecker et al.\citep{Ste06} to
the observed spectrum,
the deabsorbed spectrum seems to be too hard.

\begin{figure}
  \includegraphics[height=12pc]{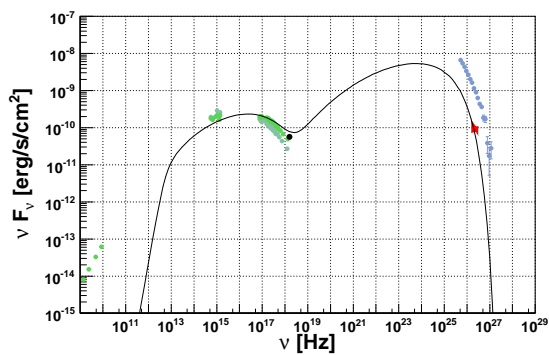}
  \caption{Spectral energy distribution of PKS\,2155-304 
in 2006. Red butterfly area indicates the deabsorbed average 
flux above 660~GeV observed with the CANGAROO-III telescope 
between July 28 and August 2, and light blue shows the results
by H.E.S.S. on July 27\citep{Aha07}. Green plots indicate 
the data of XTE and UVOT on Swift\citep{Fos07} 
and Narrabri on July 30. Deep green and black
indicate Swift data on Aug.2 and RXTE data on July 30, 
respectively.}
\label{fig:sed}
\end{figure}

Using the intrinsic energy spectrum derived above,  
the simple SSC model by Kino et al.\citep{Kin02} is applied to the
spectral energy distribution.
Figure \ref{fig:sed} shows the spectral energy distributions
of multiwavelength observations.
Adopting $5\times10^{13}\delta$~cm as the radius of the emission region
based on our results,
fitting was done to our average spectral energy distribution 
between July 28 and August 2 2006
and X-ray data taken by Swift on July 30,
which are not perfectly simultaneous observations.
We found that the beaming factor
$\delta\sim60$ and magnetic field $B\sim2.5$\,G  explain 
the observed spectral energy distribution well.
This beaming factor is much greater than 
the typical value for blazars and is consistent with
values obtained by other\,\citep{Fos07,Kus08}.
However the magnetic field is stronger compared to
those they derived.
We also try to fit the H.E.S.S. data and we found 
the smaller size of emission region and higher electron
density are required, and beaming factor
$\delta\sim80$ and magnetic field $B\sim5.0$\,G 
are obtained.
However, it is still poorly-fitting, and we need to consider 
more complicated and realistic model.

\section{conclusion}

We observed the nearby HBL PKS\,2155$-$304 in 2004, 2005,
and 2006 with the CANGAROO-III imaging atmospheric 
Cerenkov telescope.
During the active phase in July 2006, we detected a signal
at $4.8~\sigma$ level~($\sim45~\%$~Crab flux) above 660~GeV.
The intranight time variations of the flux were observed, and
obtained shortest doubling time was 34 minutes.
The size of TeV gamma-ray emission region
is limited to $5\times10^{13}\delta$~cm 
from the variability time scale, and 
a super massive black hole mass is restricted to
be less than $1.9\times10^8{\delta}\rm M_{Solar}$,
where $\delta$ is a beaming factor.
From the differential energy spectrum, an upper limit of 
the extragalactic infrared background 
light (EBL) flux is derived under the model assumption. 
The synchrotron self Compton (SSC) model fitting suggests
a larger beaming factor compared to typical blazars
and a rather larger magnetic strength.


\begin{theacknowledgments}
We thank Dr.\ W.\ Hofmann, Dr.\ S.\ Wagner, Dr.\ G.\ Rowell,
Dr.\ W.\ Benbow, Dr.\ B.\ Giebels, and L.\ Foschini
for providing details of the H.E.S.S. and the Swift
observations of PKS~2155$-$304.
This work is supported by a Grant-in-Aid for Scientific Research by the
Japan Ministry for Education, Culture, Sports, Science and Technology,
the Australian Research Council (Grants LE0238884 and DP0345983),
and Inter-University Research Program by the Institute for Cosmic Ray Research.
\end{theacknowledgments}



\bibliographystyle{aipproc}   

\bibliography{gamma08-nishijima}

\IfFileExists{\jobname.bbl}{}
 {\typeout{}
  \typeout{******************************************}
  \typeout{** Please run "bibtex \jobname" to optain}
  \typeout{** the bibliography and then re-run LaTeX}
  \typeout{** twice to fix the references!}
  \typeout{******************************************}
  \typeout{}
 }

\end{document}